\documentstyle[fleqn,12pt,epsf]{article}
\newcommand{\Pbar}{\not{\!P}}

\newcommand{\nsigma}{\mbox{\boldmath $\sigma$}}
\newcommand{\ntau}{\mbox{\boldmath $\tau$}}
\newcommand{\ngamma}{\mbox{\boldmath $\gamma$}}

\newcommand{\nh}{{\bf      h}}
\newcommand{\nj}{{\bf      j}}
\newcommand{\nk}{{\bf      k}}
\newcommand{\np}{{\bf      p}}
\newcommand{\nq}{{\bf      q}}

\newcommand{\nB}{{\bf      B}}

\begin{document}
\begin{titlepage}

\mbox{} 
\vspace*{2.5\fill} 
{\Large\bf 
\begin{center}
%****************************************************************************
%
Semi-relativistic meson exchange currents in $(e,e')$ and $(e,e'p)$ reactions
%
%*****************************************************************************
\end{center}
} 
\vspace{1\fill} 
\begin{center}
%**************************
{\large 
J.E. Amaro$    ^{1}$, 
M.B. Barbaro$  ^{2,3}$, 
J.A. Caballero$^{3}$, 
F. Kazemi Tabatabaei$ ^{1}$
}
%**************************
\end{center}
\begin{small}
\begin{center}
$^{1}${\sl 
Departamento de F\'\i sica Moderna,
Universidad de Granada, 
E-18071 Granada, SPAIN 
}\\[2mm]
$^{2}${\sl 
Dipartimento di Fisica Teorica,
Universit\`a di Torino and
INFN, Sezione di Torino \\
Via P. Giuria 1, 10125 Torino, ITALY \\[2mm]
$^{3}${\sl 
Departamento de F\'\i sica At\'omica, Molecular y Nuclear \\ 
Universidad de Sevilla, Apdo. 1065, E-41080 Sevilla, SPAIN 
}
}
\end{center}
\end{small}

\kern 1cm \hrule \kern 3mm 

\begin{small}
\noindent
%**************************************
{\bf Abstract} 
%**************************************
\vspace{3mm} 

Electron-induced one-nucleon knock-out observables are computed for
moderate to high momentum transfer making use of semi-relativistic
expressions for the one-body and two-body meson-exchange current
matrix elements.  Emphasis is placed on the semi-relativistic form of
the $\Delta$-isobar exchange current and several prescriptions for the
dynamical-equivalent form of the $\Delta$-propagator are analyzed. To
this end, the inclusive transverse response function, evaluated within
the context of the semi-relativistic approach and using different
prescriptions for the $\Delta$-propagator, is compared with the fully
relativistic calculation performed within the scheme of the
relativistic Fermi gas model. It is found that the best approximation
corresponds to using the traditional static $\Delta$-propagator.
These semi-relativistic approaches, which contain important aspects of
relativity, are implemented in a distorted wave analysis of
quasielastic $(e,e'p)$ reactions. Final state interactions are
incorporated through a phenomenological optical potential model and
relativistic kinematics is assumed when calculating the energy of the
ejected nucleon.  The results indicate that meson exchange currents
may modify substantially the $TL$ asymmetry for high missing momentum.
\kern 5mm



\end{small}

\kern 5mm \hrule \kern 1cm

\vfill

\end{titlepage}

\section{Introduction}
%======================

Electron induced one-nucleon emission reactions near the quasielastic
peak are clearly dominated by one-body (OB) dynamics since such
kinematics roughly corresponds to the virtual photon interacting
directly with a bound nucleon~\cite{Bof93,Bof96,Kel96}. Under these
conditions it has been possible to analyze a considerable amount of
data for a variety of nuclei and to study many aspects of the reaction
mechanism. Most of these studies have been based on the standard
distorted-wave impulse approximation (DWIA), where the OB current
matrix elements are computed using single-particle wave functions
obtained as solutions of the Schr\"odinger equation with
phenomenological potentials. Recently~\cite{Udi93,Udi96,Udi99,Udi01},
efforts have been made to develop a fully relativistic description of
the process in order to describe consistently the high momentum and
energy transfer region~\cite{Gao00}. This constitutes the basis of the
relativistic distorted-wave impulse approximation (RDWIA).

Within the scheme of non-relativistic approaches, the role played by
correlations beyond the mean field and their impact on the overlap
functions and spectroscopic factors has been investigated in several
works.  Many-body calculations~\cite{Van98,Maz02} including
short-range correlations (SRC) of central type only, have shown a
small effect over the quasi-hole overlap functions and, accordingly,
over the $(e,e'p)$ cross section. In~\cite{Bob94} the SRC were shown
not to modify substantially the mean field $(e,e'p)$ results at high
momentum and low excitation energy. In recent works a significant
effort has been made to improve the analysis by including the tensor
and spin-isospin channels~\cite{Mut00,Fab01} as well as long-range
correlations~\cite{MS91,MW91}.  It is important to point out that the
extraction of spectroscopic factors from the analysis of $(e,e'p)$
experiments is still not free from ambiguities. This would require an
accurate knowledge of the reaction mechanism. In this sense, a RDWIA
calculation, compared to DWIA, gives rise to a different quenching of
the $(e,e'p)$ cross section, hence leading to different spectroscopic
factors~\cite{Udi93} (larger for RDWIA).

The two-body meson exchange currents (MEC) may also play a significant
role in the description of electron scattering observables since they
are connected to the nuclear correlations by the continuity
equation. In fact the interplay between correlations and MEC in
$(e,e'p)$ reactions is far from trivial; rigorously the concept of
overlap function is not enough to describe the process in presence of
two-body current operators, and therefore the effect of correlations
in presence of MEC may be not simply parameterizable by a
spectroscopic factor.  This idea is supported by calculations of the
inclusive transverse response for nuclear matter within the correlated
basis function perturbation theory~\cite{Fab97}, that have shown a
significant effect due to MEC, contrary to the much smaller effect
usually found for uncorrelated calculations~\cite{Ama94}.  Before a
complete calculation of the exclusive response functions including MEC
within a sophisticated correlated model of the reaction be attempted,
it is necessary to calibrate different approaches to the calculation
in uncorrelated models, as a first step beyond the DWIA.

Thus in this work we restrict our attention to the mean-field
(uncorrelated) model and evaluate the role played by the MEC in
quasielastic inclusive and exclusive electron scattering reactions. In
the case of exclusive $(e,e'p)$ processes, a theoretical evaluation of
MEC, including the usual pion-in-flight (P), contact (C) and
Delta-Isobar ($\Delta$), has been presented in
refs.~\cite{Bof91,Van94,Ama99}. In~\cite{Ama99} substantial
differences were found with respect to the results of
ref.~\cite{Van94}, particularly for the interference $TL$
response. The analysis in~\cite{Van94}, which treats final state
interactions (FSI) through a real potential, was extended to higher
momentum transfer in refs.~\cite{Ryc99,Ryc01} where some relativistic
corrections were also included in the OB current operator. Instead the
standard non-relativistic MEC operators~\cite{Ris79,Ama93} were used
except for a modified ``dynamical'' $\Delta$-propagator that depends
on the invariant energy $\sqrt{s}$ of the $\Delta$.  More recently,
the theoretical model of~\cite{Bof91} has been refined in~\cite{Giu02}
modifying also the $\Delta$-propagator in order to include the
$\Delta$-invariant mass. The use of a proper ``dynamical'' propagator
in the $\Delta$-current has been discussed at length, concerning
mainly the case of two-particle emission
reactions~\cite{Wil96,Ryc96,Wil97}. One of the central questions is
how to determine $\sqrt{s}$ in a way that could be easily implemented
in existing models for $(e,e')$ and $(e,e'p)$ reactions.

In general the appropriate choice of $\sqrt{s}$ will depend on the
specific dynamical model for the reaction mechanism. Hence a proper
description of the reaction dynamics through an adequate choice of the
currents and kinematics 
%results 
is necessary. For high momentum transfer
relativistic effects play a significant role and therefore relativity
should enter in the description of kinematics and/or current matrix
elements.  Within a fully relativistic model, MEC effects in $(e,e'p)$
have been computed recently in~\cite{Meu02}, where only the contact
(or ``seagull'') current has been considered. Due to the complexity of
the fully relativistic two-body currents, particularly the $\Delta$
one, in this work we start by developing reasonable semi-relativistic
expressions for all the two-body currents (P, C and $\Delta$) that can
be easily implemented in existing non-relativistic descriptions of the
reaction mechanism.  These semi-relativistic currents retain important
aspects of relativity and hence they may be used to describe properly
recent experiments performed at high momentum and energy transfers.

In this paper we discuss the one-particle emission sector, introducing
a recently developed semi-relativistic (SR) approximation for the
three MEC operators which, in conjunction with the SR form of the OB
current derived in~\cite{Ama96a} (see~\cite{Rep02} for a recent review
on these expansions), makes it possible to evaluate quasielastic
$(e,e')$ and $(e,e'p)$ observables.  The SR-OB current was applied in
inclusive $(e,e')$ \cite{Ama96b}, and for coincidence $(e,e'p)$
~\cite{Udi99,Maz02,Ama98a,Ama99b,Ama02a} reactions. The SR expressions
for the MEC operators were obtained in~\cite{Ama98b} in the case of
the pionic and contact currents. In~\cite{Rep02,Ama02b} a comparison
between the SR-MEC predictions for the inclusive transverse response
and the exact RFG results is presented.

One of the goals of this work is to derive a semi-relativistic
expression for the $\Delta$-exchange current and compare its
prediction for the inclusive transverse response function with the
exact relativistic result obtained within the RFG model of
ref.~\cite{Ama03}.  This comparison, carried out in the inclusive
channel (section~2), makes it possible to test the reliability of the
different prescriptions of the ``dynamical'' $\Delta$-propagator to be
used in exclusive $(e,e'p)$ processes (section~3). Once this is
settled, the impact of the relativistic MEC over the exclusive
response functions is evaluated for moderate to high momentum
transfer. Finally, in section~4 we draw our conclusions.

%======================================

\section{Inclusive transverse response}

%======================================

%----------------
%    Figure 1
%----------------
\begin{figure}[tb]
\begin{center}
\leavevmode
% scala la figura di  un fattore 0.9:
\def\epsfsize#1#2{0.9#1}
\epsfbox[200 450 400 720]{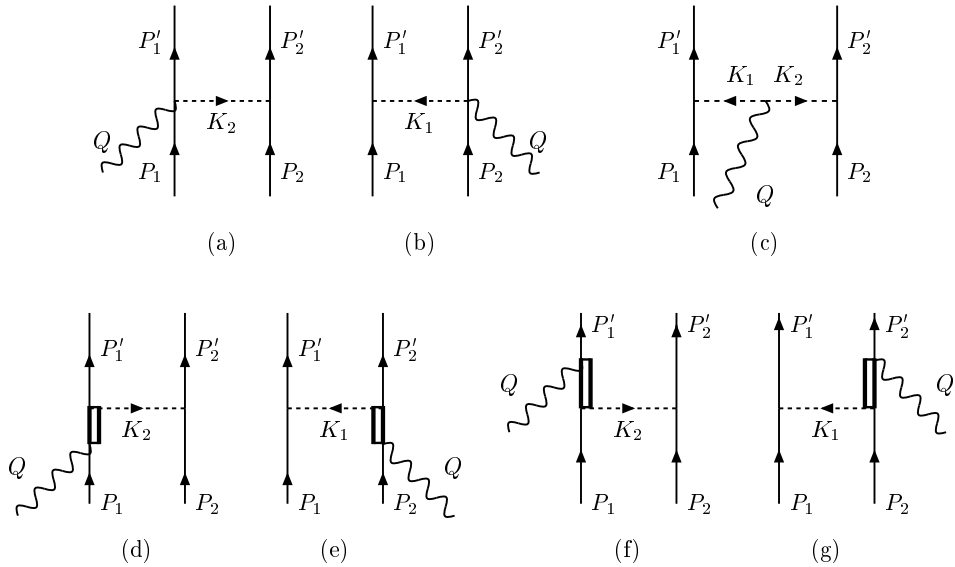}
\end{center}
\caption{Feynman diagrams contributing to the two-body current with
one pion-exchange. Contact (C) (a,b), pionic (P) (c), and isobar
($\Delta$) (d)--(g) are included in this work.}
\end{figure}

The impact of meson exchange currents on inclusive electron scattering
has been traditionally investigated in a non-relativistic framework
and relativistic corrections have been implemented making use of a
power expansion in the momenta of the nucleons and of the exchanged
photon~\cite{Van81,Alb90,Alb93}.  In refs.~\cite{Rep02,Ama03} a
semi-relativistic analysis of the MEC (P, S and $\Delta$) currents,
represented in fig.~1, has been performed in the context of the RFG
with the aim of finding simple prescriptions for implementing
relativistic effects into non-relativistic calculations.  Within the
RFG model, it is possible to perform a fully relativistic calculation
of the inclusive response functions, using exact relativistic
kinematics, currents and propagators.  Hence the inclusive transverse
response function of nuclear matter can be used to test the quality of
the semi-relativistic MEC, to be applied in the next section to
semi-inclusive processes.

In this section we focus on the $\Delta$ current, represented by
diagrams (d--g) of fig.~1, putting emphasis on the ``dynamical''
treatment of the $\Delta$ propagator, an aspect of the problem which
was neglected in the SR model of~\cite{Ama03}.  Restricting our
attention to the case of one-particle emission processes, the
contribution of the $\Delta$ current to the transverse nuclear
response function is obtained as the interference between the $\Delta$
and the OB currents. In the Fermi gas model it reads
\begin{eqnarray}
R^T_{\Delta\mbox{-OB}}(q,\omega)
&=& \frac{3Z}{8\pi k_F^3} 2{\rm Re} \int_{h<k_F} d^3 h 
\delta(E_{\np}-E_{\nh}-\omega)
{\cal N}_{\np\nh}
{\rm Tr} \left[ \nj^{OB}_T(\np,\nh)^* \cdot\nj^{\Delta}_T(\np,\nh)
  \right] \, ,
\nonumber\\
\end{eqnarray}
where $\np=\nh+\nq$ is the momentum of the ejected particle, $\nj_T$
is the transverse (perpendicular to $\nq$) component of $\nj$ and
$k_F$ is the Fermi momentum.  In the RFG model
$E_\nk=\sqrt{\nk^2+m_N^2}$, whereas in the non-relativistic (NR) Fermi
gas, $E_\nk=m_N+\epsilon_{\nk}=m_N+\nk^2/(2m_N)$.  Moreover, the
factor ${\cal N}_{\np\nh}$, arising from the spinor normalization, is
$m_N^2/(E_{\np}E_{\nh})$ for RFG and 1 in a NR model.  
Finally $\nj^a_T(\np,\nh)$ (which implicitly include
spin and isospin indices) are the  OB or $\Delta$ particle-hole  
currents \cite{Rep02}.

The non relativistic two-body $\Delta$ current used in this paper has
been obtained from the relativistic one in the limit of small momenta
by following the procedure sketched in ref.~\cite{Ama03}; it can
be written in momentum space as
\begin{eqnarray}
\lefteqn{\nj^\Delta_{nr}(\np'_1,\np'_2,\np_1,\np_2)}
\nonumber\\
&=&
\frac{1}{9} 
\frac{G_1}{2m_N}\frac{f_{\pi N \Delta}}{m_\pi}\frac{f}{m_\pi}
\left\{
G_\Delta(P_1+Q)
\nq\times 
\left[ -\nk_2\times\nsigma^{(1)} +2i\nk_2 \right]
\left[ 2\tau_3^{(2)}-i[\ntau^{(1)}\times\ntau^{(2)}]_z \right]
\right.
\nonumber\\
&&
\mbox{} +
\left.
G_\Delta(P'_1-Q)
\nq\times 
\left[ \nk_2\times\nsigma^{(1)} +2i\nk_2 \right]
\left[ 2\tau_3^{(2)}+i[\ntau^{(1)}\times\ntau^{(2)}]_z \right]
\right\}
\frac{\nk_2\cdot\nsigma^{(2)}}{\nk_2^2+m_\pi^2}
\nonumber\\
&& 
\mbox{} +
(1\longleftrightarrow 2)\,,
\label{nrcurrent}
\end{eqnarray}
where $P_i$, $P'_i$ are the four-momenta of the initial and final
nucleons defined in fig.~1, $\nk_i=\np'_i-\np_i$ is the momentum
transferred to the $i$-th nucleon and $G_\Delta(P)$ is the
non-relativistic (dynamical) $\Delta$ propagator, which will be later
discussed in detail. We use the following values for the coupling
constants: $G_1=4.2$, $f^2/4\pi = 0.079$, $f_{\pi N \Delta} = 2.24$.

In \cite{Ama03} a more general relativistic current was considered,
containing additional couplings $G_2$, $G_3$ and dependence on the
so-called off-shell parameters $z_1,z_2,z_3$. Therein it was shown
that the terms $G_2$ and $G_3$ have a weak impact on the transverse
response for $q$ below 1 GeV and that this response displays a small
dependence upon the off-shell parameters compared with its sensitivity
to $\sqrt{s}$, which will be discussed at the end of this Section.
Therefore in this work we use the Peccei lagrangian, which neglects
$G_2$ and $G_3$ and corresponds to $z_i=-1/4$.

%----------------
%    Figure 2
%----------------
\begin{figure}[t]
\begin{center}
\leavevmode
% scala la figura di  un fattore 0.9:
\def\epsfsize#1#2{0.9#1}
\epsfbox[200 450 400 720]{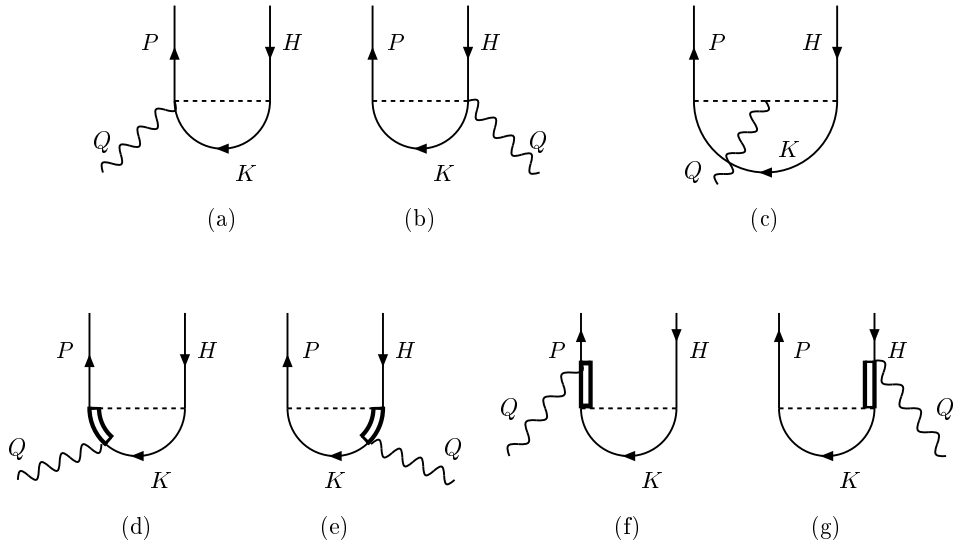}
\end{center}
\caption{Exchange MEC diagrams contributing to the one-particle
emission responses computed in this work.  In the vertex-type isobar
diagrams (d,e) different prescriptions for the dynamical $\Delta$
propagator are discussed in the text.}
\end{figure}

The MEC ph matrix elements are obtained by summing over all the
single-particle states occupied in the ground-state Slater
determinant \cite{Rep02,Ama03}.
In the case of nuclear matter only the exchange term, represented by
the many-body diagrams of fig.~2, survives, due to spin and isospin
symmetries.  Moreover, it can be proved that the exchange $\Delta$
terms represented by diagrams (f) and (g) are also zero in nuclear
matter.  In the case of finite nuclei there exists however a
contribution coming from the direct $\Delta$ term, which, although
small, has been included in the calculations presented in next
section.

Using eq.~(\ref{nrcurrent}) the particle-hole
matrix element of the non-relativistic $\Delta$ current in nuclear
matter results
\begin{eqnarray}
\nj^\Delta_{nr}(\np,\nh)
&=&
 -i\delta_{t_pt_h}\tau_h\frac{4}{9}
\frac{G_1}{2m_N}\frac{f_{\pi N \Delta}}{m_\pi}\frac{f}{m_\pi}
\int  \frac{d^3 k}{(2\pi)^3} \theta(k_F-k)
\nonumber\\
&& 
\mbox{}\times
\left\{ 
G_{\Delta}(\nk+\nq)\nB(\nh-\nk)_{s_ps_h}
+G_{\Delta}(\nk-\nq)\nB(\np-\nk)_{s_ps_h}
\right\} \, ,
\label{nr}
\end{eqnarray}
where $\tau_h=1$ if $h$ is a proton and $-1$ if  neutron, and
we have defined the function
\begin{equation}
\nB(\np-\nk)_{ss'}=
\frac{[\nsigma_{ss'}\cdot(\np-\nk)](\np-\nk)+(\np-\nk)^2\nsigma_{ss'}}
{(\np-\nk)^2+m_\pi^2}\ .
\end{equation}
If the $\nk$-dependence of the $\Delta$-propagator is neglected, then
the integral in eq.~(\ref{nr}) can be solved
analytically\footnote{Note that in this section we are not including
$\pi$N form factors in order to obtain an analytical result for
numerical convenience.  In the next section they will be properly
included.}  (see ref.~\cite{Ama94} for its explicit form).

The relativistic expression for the currents can be found
in~\cite{Ama03} and yields results which differ significantly from the
non-relativistic ones even if the momentum transfer is as low as 500
MeV/c.  This difference is due in part to the relativistic kinematics
(RK), which can be easily implemented in the NR model by the
replacement $\omega\rightarrow \omega(1+\omega/2m_N)$ \cite{Ama96b}, with the
exception of the electromagnetic form factors, that must be computed
for the unshifted value of $\omega$.  Moreover, the NR results can be
brought closer to the relativistic ones if the SR form of the current
operators
\begin{equation}
\nj^a_{T,SR}(\np,\nh)
\equiv \frac{1}{\sqrt{1+\tau}}\nj^{a}_{T,nr}(\np,\nh) 
\label{SROB}
\end{equation}
is employed for $a=$OB,C,P,$\Delta$, with $\tau=|Q^2|/4m_N^2$.

Eq.~(\ref{SROB}) has been obtained 
(see refs.~\cite{Rep02,Ama98b,Ama03} for details) by a direct Pauli
reduction, expanding in powers of $\nh/m_N$ to first order 
the OB current~\cite{Ama96a} and to leading order 
the MEC of pionic and contact types~\cite{Ama98b,Ama02b}.  
In this paper we use a similar `factorized'
expression for the SR $\Delta$-current that was already proposed
in~\cite{Ama03}.  The $\tau$-dependent factor 
%($\tau=|Q^2|/4m_N^2$)
in eq.~(\ref{SROB}) arises from the spinology and produces a reduction
of the responses.

Let us now discuss the $\Delta$ propagator $G^\Delta$, appearing in
the $\Delta$ current, eqs.~(\ref{nrcurrent},\ref{nr}).  In a fully
relativistic theory, $G^\Delta$ is given by the Rarita-Schwinger
tensor~\cite{Ama03}. The possible decay of the isobar into a $N\pi$
state is accounted for by the substitution of the $\Delta$ mass
$m_\Delta \rightarrow m_\Delta-\frac{i}{2}\Gamma(P^2)$ in the
denominator of the $\Delta$ propagator, where the function
$\Gamma(P^2)$ is the width of the resonance (see
\cite{Nie93a,Nie93b}).

The non-relativistic version of this propagator is 
defined by the positive energy sector 
for spatial indices and small momenta \cite{Ama03}
\begin{equation}
G^\Delta_{ij}(P) 
\simeq
\frac{ \Pbar+m_\Delta}{P^2-m_\Delta^2}
\left( \delta_{ij} + \frac{1}{3} \gamma_i\gamma_j \right) 
\longrightarrow
G_{\Delta}(P)
\left( \frac23\delta_{ij} -
\frac{i}{3}\epsilon_{ijk}\sigma_k\right) \, .
\label{nrpropagator}
\end{equation}
Note that diagrams (d--g) of fig.~1 correspond to different momenta in
the propagators, namely $G_\Delta(P_1+Q)$ and $G_\Delta(P'_1-Q)$:
these, referred to as $\Delta$-excitation ($G_\Delta^I$) and
$\Delta$-deexcitation ($G_\Delta^{II}$) in \cite{Ryc99,Giu02}, denote
the propagator for a $\Delta$ created after and before
photo-absorption, respectively.

The static limit approximation to $G^\Delta$ consists in taking
\begin{eqnarray}
\frac{ \Pbar+m_\Delta}{P^2-m_\Delta^2}
\simeq
\frac{ m_N\gamma_0+m_\Delta}{P^2-m_\Delta^2}
\rightarrow \frac{ m_N+m_\Delta}{m_N^2-m_\Delta^2}
=\frac{1}{m_N-m_\Delta} \, ,
\label{static}
\end{eqnarray}
where we have used $P^2\simeq m_N^2$ for $P=P_1+Q$ or $P=P'_1-Q$ and
$q$, $\omega$ small, since $P_1$ and $P'_1$ are the four-momenta of
the initial and final nucleons. In this case, the two propagators are
equal, i.e., $G_\Delta(P_1+Q)=G_\Delta(P'_1-Q)=(m_N-m_\Delta)^{-1}$
and can be factorized out in eq.~(\ref{nrcurrent}), thus yielding the
traditional form of the non-relativistic current~\cite{Ama94,Ris79}.

The effects introduced by relativity can be appreciated in
fig.~3. Here we show the inclusive transverse response of $^{40}$Ca
for two values of the momentum transfer, $q=0.5$ GeV/c (top panel) and
$q=1$ GeV/c (bottom panel).  We use Fermi momentum $k_F=237$ MeV/c and
the Galster parameterization of the nucleon form factors \cite{Gal71}.
For the $\Delta$ current we use the electric form factor of the
proton.  For the comparison with the non-relativistic result we use
strong form factors $F_{\pi NN}=F_{\pi N\Delta}=1$.  The fully
relativistic calculation performed within the RFG model of
ref.~\cite{Ama03} (thick solid line) is compared with various
non-relativistic approaches: the traditional non-relativistic model
(NR), the NR model but including relativistic kinematics (denoted as
RK), and the results corresponding to the semi-relativistic currents,
using in addition relativistic kinematics (denoted as SR0).  Apart
from the RFG, where the Rarita-Schwinger propagator is used including
the $\Delta$ width, the other results have been evaluated employing
the static limit approximation for the $\Delta$ propagator, for which
$\sqrt{s}=m_N$, and therefore $\Gamma_\Delta(s)=0$, since $s$ is below
the threshold, $m_N+m_\pi$. Note the crucial role played by the
relativistic kinematics and, moreover, how the SR approach, compared
to the RK case, improves the agreement with the RFG results.

%----------------
%    Figure 3
%----------------
\begin{figure}[t]
\begin{center}
\leavevmode
% scala la figura di  un fattore 0.9:
\def\epsfsize#1#2{0.8#1}
\epsfbox[70 400 470 790]{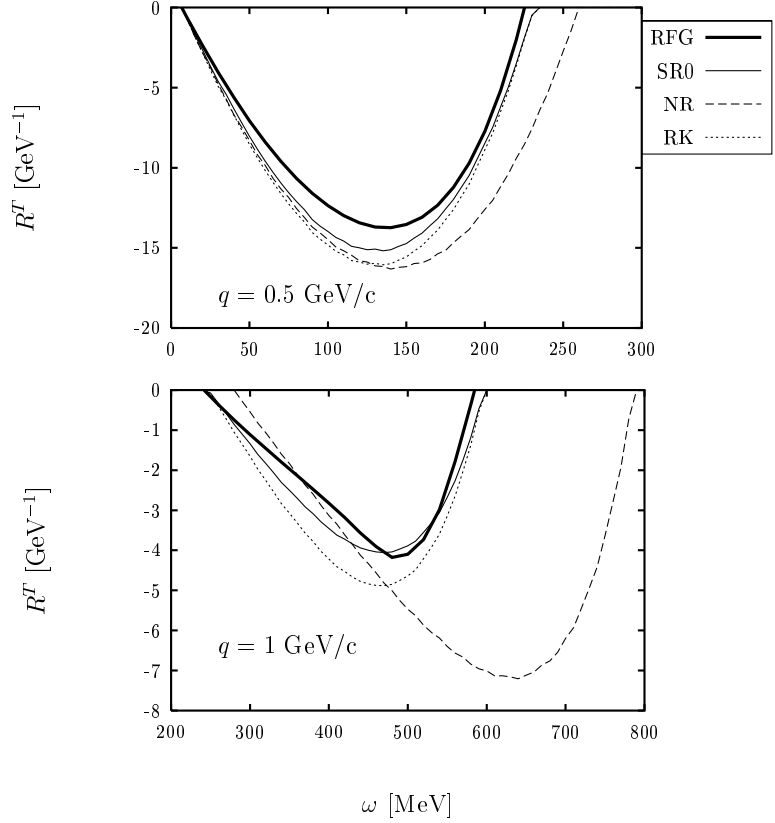}
\end{center}
\caption{Contribution of the $\Delta$ to the transverse response
function for $q=500$ and 1000 MeV/c.  The results corresponding to the
relativistic Fermi gas (RFG) are compared with the non-relativistic
(NR) Fermi gas, including relativistic kinematic (RK), and using the
semi-relativistic (SR0) approach for the electromagnetic currents.  }
\end{figure}

Next we study the effect of including a ``dynamical'' $\Delta$
propagator within the SR model. This can be done by using different
approximations for the propagators, $G_\Delta (K+Q)$ and $G_\Delta
(K-Q)$, appearing in the non-relativistic current, eq.~(\ref{nr}),
under the requirement that the new ``dynamical'' propagator be
independent of $\nk$: only in this case it can be easily introduced
into non-relativistic calculations.

The results for the $\Delta$-OB transverse response computed within
the SR model, and using different prescriptions for the $\Delta$
propagator, are displayed in fig.~4.  As in the previous figure and
for comparison, we present the fully relativistic results (RFG) and
the SR approach using the static limit for the $\Delta$ propagator
(SR0). In all the cases, relativistic kinematics is assumed.  The
various approaches considered to describe the ``dynamical'' $\Delta$
propagator are below listed and discussed in some detail.
\begin{itemize}
\item[1)]
The propagator is written, as suggested in refs.~\cite{Wil96,Wil97}, as
\begin{equation}
G_\Delta(P) = \frac{1}{\sqrt{s}-m_\Delta + \frac{i}{2}\Gamma_{\Delta}(s)}\ ,
\label{propSR1}
\end{equation}
where $s=P^2$ and $\sqrt{s}$ is the available energy in the $\Delta$
rest system.  The approximate values of $s$ are obtained by neglecting
the momentum $k$ and kinetic energy $\epsilon_\nk$ compared with $q$ and
the nucleon mass $m_N$, i.e.,
\begin{eqnarray}
s^I &\equiv& (K+Q)^2 
= (E_{\nk}+\omega)^2-(\nk+\nq)^2 \simeq (m_N+\omega)^2-q^2
\label{sI}
\\
s^{II} &\equiv& (K-Q)^2 \simeq  (m_N-\omega)^2-q^2 \, .
\label{sII}
\end{eqnarray}
As we observe in fig.~4, this prescription (denoted as SR1) produces a
hardening of the response function, that is more pronounced for high
$q$, and under-estimates the exact result at the peak.

\item[2)] This second prescription is obtained, as suggested in
refs.~\cite{Giu02,Wil97}, by using the static approximation in the
de-excitation diagram, fig.~2e, i.e., $s^{II}\simeq m_N^2$, and a
dynamical propagator in the $\Delta$ excitation diagram, fig.~2d. In
this case we use the same value $s^I$ as in eq.~(\ref{sI}). As shown
in fig.~4, this prescription (labelled as SR2) still produces a
hardening of the response, but now the response function is larger
than the relativistic one. We note also that the SR2 curve crosses the
static SR0 result precisely at the position of the quasielastic peak
(QEP). In fact, this is determined by $(m_N+\omega)^2 -q^2= m_N^2$,
hence eq.~(\ref{sI}) gives the static result, $\sqrt{s^I}=m_N$.

\item[3)]
From eq.~(\ref{nrpropagator}) we may write
\begin{eqnarray}
\frac{ \Pbar+m_\Delta}{P^2-m_\Delta^2}
\simeq
\frac{E\gamma_0 -\nq\cdot\ngamma + m_\Delta}{\sqrt{s}+m_\Delta}
\frac{1}{\sqrt{s}-m_\Delta}
\label{residuo}
\end{eqnarray}
for $P=K+Q$. The result given by eq.~(\ref{propSR1}) is re-obtained
when the first term in the right hand side of eq.~(\ref{residuo}) is
close to one. Note that for $q$ large, $\sqrt{s}$ can be significantly
different from $E$, while one cannot neglect the term
$\nq\cdot\ngamma$.  This would give an additional spin-dependent term
and the propagator would get tangled. Instead, a possible way to
proceed is by taking partially into account the $q$-dependence in the
numerator in the way:
\begin{eqnarray}
G_\Delta(K+Q)
&\simeq& 
\frac{m_N+\omega+q + m_\Delta}{\sqrt{s^I}+m_\Delta}
\frac{1}{\sqrt{s^I}-m_\Delta}
\\
G_\Delta(K-Q)
&\simeq& 
\frac{m_N+\omega-q + m_\Delta}{\sqrt{s^{II}}+m_\Delta}
\frac{1}{\sqrt{s^{II}}-m_\Delta} \, ,
\end{eqnarray}
where $s^{I,II}$ are given in eqs.~(\ref{sI},\ref{sII}). This
procedure allows us to write down an approximate diagonal expression
for the $\Delta$ propagator, and the corresponding results are denoted
in fig.~4 as SR3. The reduction of the response function is shown not
to be as large as in the SR1 case.

\end{itemize}

%----------------
%    Figure 4
%----------------
\begin{figure}[t]
\begin{center}
\leavevmode
% scala la figura di  un fattore 0.9:
\def\epsfsize#1#2{0.8#1}
\epsfbox[70 400 470 790]{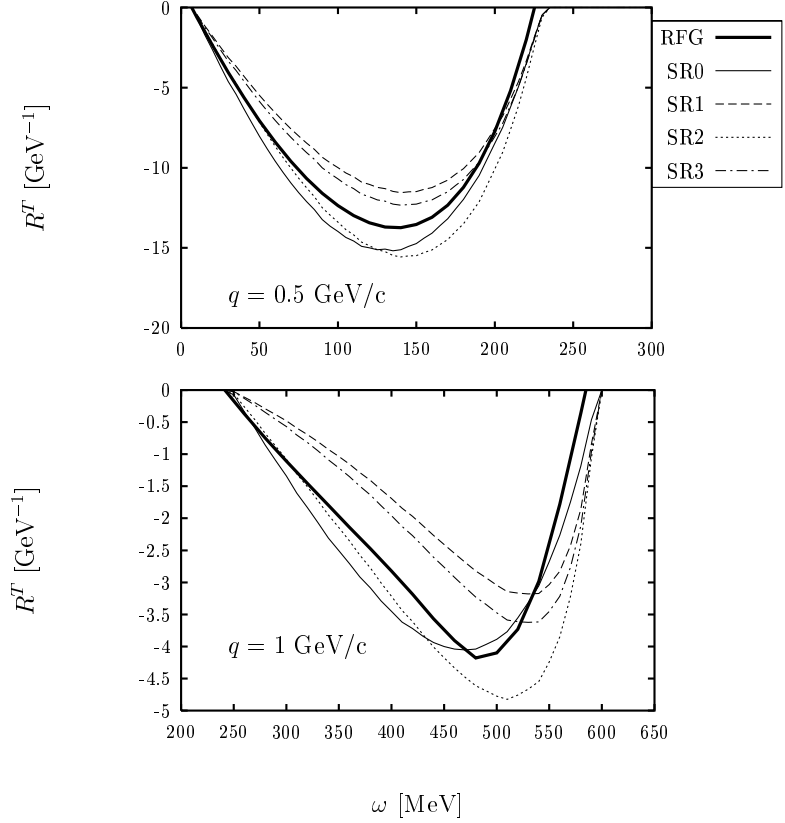}
\end{center}
\caption{Contribution of the $\Delta$ to the transverse response
function.  using different prescriptions for the $\Delta$ propagator,
discussed in the text }
\end{figure}

Summarizing, from fig.~4 we may conclude that none of the above
approximations proposed for the dynamical $\Delta$ propagator is
entirely satisfactory. In fact we observe that the best result,
compared with the fully relativistic RFG one, corresponds to the
static limit approximation.

To complete this discussion, we note that the role of the width
$\Gamma$ in the SR models of fig.~4 is irrelevant since the
approximate values taken for $\sqrt{s}$ are far below the pole for
these kinematics.  In fact, for the prescription SR1 we have from
eq.~(\ref{sI}):
\begin{equation}
\omega = \sqrt{s^I+ q^2} - m_N \, .
\label{nrpole}
\end{equation}
This means that for
$q=1$ GeV/c and $\sqrt{s^I}=m_\Delta$, we get $\omega\sim649$
MeV, which is well above the allowed region of the QEP, while the 
threshold, $\sqrt{s^I}=m_N+m_\pi$, is reached for $\omega\sim531$ MeV,
below which the $\Delta$ width is zero. Hence only the small tail of
the $\Delta$ close to the threshold is being considered in this region
of energy.

Actually in the relativistic model things are totally different
because the pole is reached inside the allowed energy region.  Indeed
the delicate energy balance in the denominator makes the value
of the inner momentum $\nk$ to play a role. 
Instead of (\ref{nrpole}) a better approximation to the exact relation
is represented by
\begin{equation}
\omega = \sqrt{s^I+ (\nq+\nk)^2} - m_N \, .
\label{rpole}
\end{equation}
In this case the pole is first reached at
\begin{equation}
\omega = \sqrt{m_\Delta^2+ (q-k_F)^2} - m_N \, , 
\label{firstreached}
\end{equation}
which gives $\omega\sim 511$ MeV, close to the peak for $q=1$ GeV/c. 

Thus we have considered two new prescriptions, denoted as SR4 and
SR5. To help the reader they are shown in a separate figure (fig.~5),
where they are again compared with the fully relativistic calculation
(RFG) and the static limit approach (SR0). These new prescriptions are
based on the value assigned to $\sqrt{s}$, such that the position of
the pole be closer to the relativistic case. More precisely:

\begin{itemize}
\item[4)] One considers $\sqrt{s^I}=\sqrt{s_{NN}}-m_N$, being $s_{NN}$
the invariant energy of the two outgoing nucleons, suggested
in~\cite{Wil96,Wil97} for two-nucleon knock-out, and apparently
applied also to one-nucleon emission in~\cite{Giu02}. The meaning of
$s_{NN}$ for this case is doubtful, since there is only one particle
in the final state. However, from the $\Delta$-excitation diagram,
fig.~2d, we see that $\nk,\nh$ are two entering momenta and $\np,\nk$
the two exiting ones.  Hence the prescription SR4 in fig.~5 is based
on the approximation:
\begin{equation}  
s_{NN} \simeq (P+K)^2  = (E_{\nh}+E_{\nk}+\omega)^2-(\nh+\nk+\nq)^2
\simeq (2m_N+\omega)^2-q^2 \ .
\end{equation} 
In this case the pole for $q=1$ GeV/c is reached at $\omega\simeq 513$
MeV, inside the QE region.  However, from the results in fig.~5, this
prescription appears to be less satisfactory than the previous ones.

\item[5)]
Finally, the prescription SR5 is obtained by exploiting
eq.~(\ref{firstreached})
that gives the right position at which  the pole is first reached
and defining 
\begin{eqnarray}
s^I &=& (m_N+\omega)^2-(q-k_F)^2
\\
s^{II} &=& (m_N-\omega)^2-(q-k_F)^2 \ .
\end{eqnarray}
The results show that for $q=1$ GeV/c the pole is reached at the right
position close to the peak of the response. The change of sign of the
response is a consequence of the change of sign of the $\Delta$
propagator when one crosses the pole.
\end{itemize}

%----------------
%    Figure 5
%----------------
\begin{figure}[t]
\begin{center}
\leavevmode
% scala la figura di  un fattore 0.9:
\def\epsfsize#1#2{0.8#1}
\epsfbox[70 400 470 790]{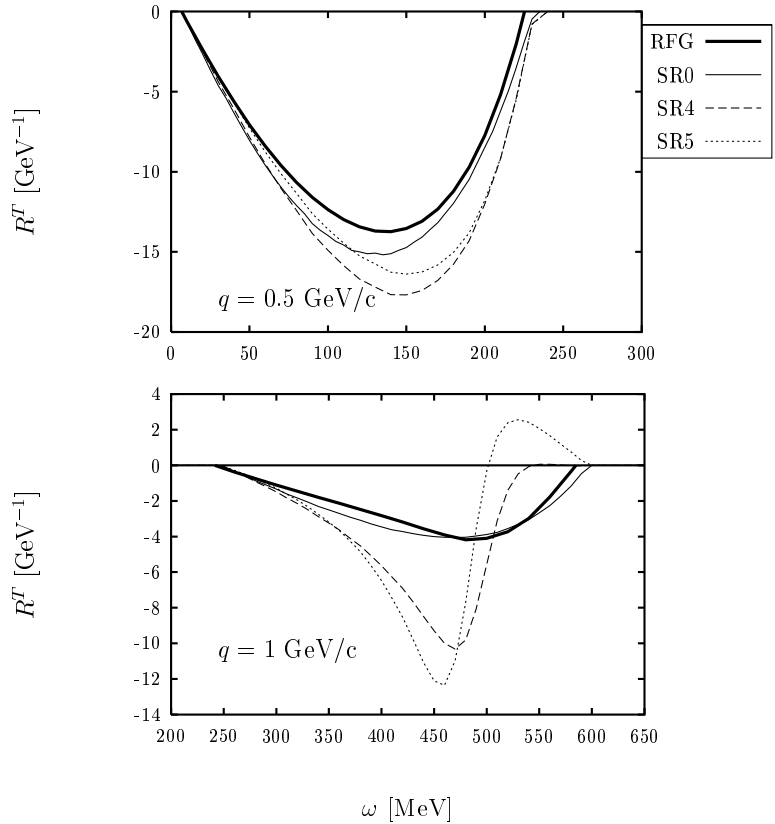}
\end{center}
\caption{ Contribution of the $\Delta$ to the transverse response
function using different prescriptions for the $\Delta$ propagator
discussed in the text.  }
\end{figure}

From fig.~5 we observe that it is not possible to recover the exact
relativistic result by using a $\Delta$ propagator independent of
$\nk$, even if it reaches the pole (in absence, of course, of the
$\Delta$ width) inside the allowed region. In fact the present
singularity can only be smoothed, even in the presence of the $\Delta$
width, after an integration over $\nk$ is performed, as in the
relativistic model.

In general we conclude that, if a dynamical propagator has to be used,
like in the semi-relativistic approaches discussed here, a smoothed
form not hitting the pole is needed in order not to deviate too much
from the exact result for the response function. Among the
prescriptions analyzed in this section, the static form of the
$\Delta$ propagator produces, in spite of its simplicity, the best
agreement with the RFG results in the case of one-particle knockout
sector, once relativistic kinematics and semi-relativistic corrections
of the currents are used. Therefore, a procedure to ``dynamize'' the
$\Delta$ propagator as the ones presented here, and its use within the
context of a non-relativistic or semi-relativistic distorted wave
analysis of quasielastic $(e,e')$ and $(e,e'p)$ reactions appears not
to give better results than the static approximation.  Obviously, this
conclusion does not affect the validity of the approximations
discussed in~\cite{Wil97} for the two-particle emission channel, where
the kinematical conditions are completely different.

In the next section we apply the present semi-relativistic model of
MEC to the exclusive $(e,e'p)$ response functions of nuclei.

%==============================
\section{Exclusive $(e,e'p)$ observables}
%==============================

In this section we compute the quasielastic $(e,e'p)$ response
functions for intermediate to high momentum transfer values within the
context of the SR approach introduced in the previous section. As
shown in the case of quasielastic inclusive $(e,e')$
responses~\cite{Ama96a,Rep02,Ama02b,Ama03}, accounting for
relativistic effects requires at least to treat properly the
kinematics and relativistic factors in the currents. Referring to the
$\Delta$ current, we make use of the static limit approach for the
$\Delta$ propagator, as this gives rise to the best agreement with the
fully-relativistic calculation in the inclusive channel, as shown in
the previous Section.  As a complete relativistic analysis of all the
MEC (P, C and $\Delta$) currents in $(e,e'p)$ processes is still
lacking, the use of ``dynamized'' $\Delta$ propagators within existing
NR or SR descriptions of the reaction mechanism does not appear to be
well founded and, moreover, these may lead to large discrepancies with
the exact calculation.

The general formalism for $(e,e'p)$ reactions has been presented in
detail in refs.~\cite{Bof96,Kel96,Udi93,Ras89} and we refer to them
for specifics on the kinematics. Assuming plane waves for the incoming
and outgoing electron (treated in the extreme relativistic limit) and
parity conservation, the exclusive cross section can be written in the
form
\begin{equation}
\frac{d^5\sigma}{d\epsilon' d\Omega'_e d\Omega_{p'}}=
K \sigma_M 
\left( {v}_L W^L + {v}_T W^T + {v}_{TL} W^{TL}\cos\phi' 
     + {v}_{TT} W^{TT} \cos 2\phi'
\right) \, ,
\label{exccs}
\end{equation}
where $\epsilon'$ and $\Omega'_e$ are the energy and solid angle
corresponding to the scattered electron and
$\Omega_{p'}=(\theta',\phi')$ is the solid angle for the ejected
proton with four-momentum $P'^\mu=(E',\np')$. In eq.~(\ref{exccs})
$K=2p'm_N/(2\pi\hbar)^3$ and $\sigma_M$ is the Mott cross section.
Finally, ${v}_\alpha$ are the electron kinematical factors given in
refs.~\cite{Ama96b,Ras89}.  The labels $L$ and $T$ refer to the
longitudinal and transverse projections of the current matrix elements
with respect to the virtual photon direction, respectively.

The hadronic content of the problem enters via the response functions
$W^\alpha$, which are obtained by taking the appropriate components of
the hadronic tensor
\begin{equation}
W^{\mu\nu}=\frac{1}{K}
\sum_{m_s M_\alpha}
\left\langle 
\np' m_s, \Phi_{\alpha}^{(A-1)}|\hat{J}^\mu(Q)|\Phi_0^{(A)}
\right\rangle^*
\left\langle 
\np' m_s, \Phi_{\alpha}^{(A-1)}|\hat{J}^\nu(Q)|\Phi_0^{(A)}
\right\rangle \, ,
\label{wmunu}
\end{equation}
where a sum over undetected final polarization states is performed. In
eq.~(\ref{wmunu}), $\hat{J}^\mu (Q)$ is the nuclear current operator
and we assume the initial state $|\Phi_0^A\rangle$ to correspond to a
spin-zero nuclear target in its ground state with energy $E_0^{(A)}$.
The final state $|\np' m_s, \Phi_{\alpha}^{(A-1)}\rangle$ is assumed
to behave asymptotically as a knockout nucleon with momentum $\np'$
and spin quantum number $m_s$ and a residual nucleus left in a bound
state, i.e., $|\Phi_\alpha^{(A-1)}\rangle=|J_\alpha,M_\alpha\rangle$,
with energy $E_\alpha^{(A-1)}$.

In the present distorted wave analysis of $(e,e'p)$ reactions, the
outgoing nucleon state is described by a wave function solution of the
Schr\"odinger equation with an optical potential $V_{opt}$ fitted to
elastic nucleon-nucleus scattering data.  The matrix elements of the
hadronic tensor are computed by performing a multipole expansion of
both the distorted nucleon wave and the current operators in terms of
the usual Coulomb $C_J$, electric $E_J$ and magnetic $M_J$ multipoles
(see refs.~\cite{Maz02,Ama96b,Ama98a} for details on the model,
and~\cite{Ama94,Ama93,Ama96b} for explicit expressions of the
multipole matrix elements of the currents).

The electromagnetic form factors used in the contact and pionic MEC
are given in Ref.~\cite{Ama93}, while for the $\Delta$ we use the same
as the electromagnetic proton form factor.  In this section we include
monopole form factors in the MEC $F_{\pi NN}(K)=F_{\pi
N\Delta}(K)=(\Lambda^2-m_\pi^2)/(\Lambda^2-K^2)$ with $\Lambda=1300$
MeV.  For low $q$ these form factors are small~\cite{Ama93}, but for
high values of $q$ (close to 1 GeV, as here) they can reduce the MEC
contribution of about 25\%.

Note that the $\Delta$-current contributions can also be sensitive to
the $\Delta$-nucleus potential in the medium (see Ref.~\cite{Ryc99}).
Indeed in-medium modifications of the $\Delta$ and $\pi$ propagators
have been studied in inclusive $(e,e')$ scattering in the region of
the $\Delta$ peak and for pion electro-production \cite{Gil97}, but
not, to our knowledge, for the MEC in the QEP.  A rough estimate of
these effects can be performed by adding a constant $\Delta$-nucleus
optical potential $V_{\Delta}=-30 -40 i $ MeV~\cite{Che88} in the
denominator of the static $\Delta$ propagator: since $V_\Delta$ turns
out to affect our results very little, we have accordingly neglected
it.

%----------------
%    Figure 6
%----------------
\begin{figure}[tp]
\begin{center}
\leavevmode
% scala la figura di  un fattore 0.9:
\def\epsfsize#1#2{0.8#1}
\epsfbox[50 280 500 800]{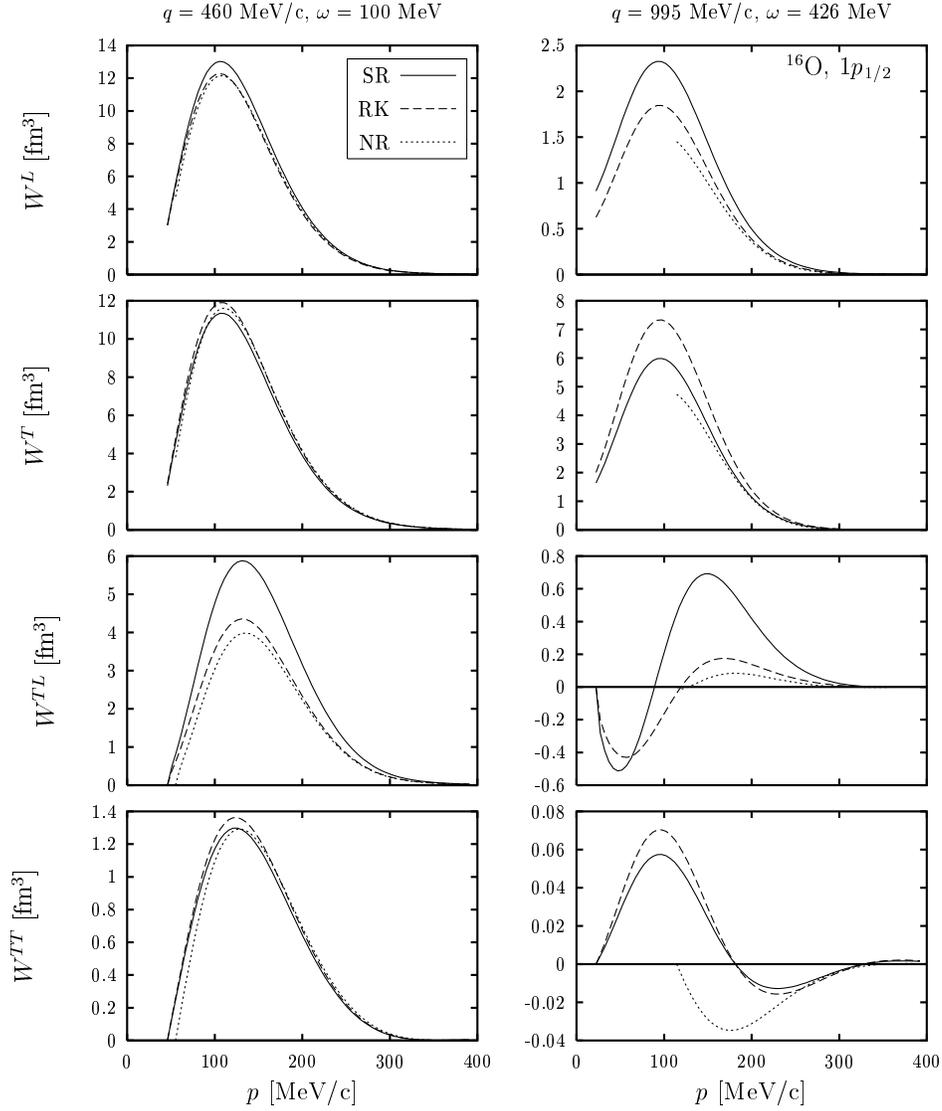}
\end{center}
\caption{ Exclusive response functions for proton knock-out from the
$1p_{1/2}$-shell in $^{16}$O. Two $(q,\omega)$-constant kinematical
situations are selected. The traditional non-relativistic results (NR)
are compared with the responses obtained including relativistic
kinematics (RK) and using the semi-relativistic (SR) form of the
currents.  All the curves include MEC.  }
\end{figure}

To illustrate the role of the relativistic corrections we display in
fig.~6 the four unpolarized exclusive response functions for proton
knock-out from the $1p_{1/2}$-shell in $^{16}$O leading to the
residual nucleus $^{15}$N. The kinematics selected corresponds to
$(q,\omega)$-constant kinematics (sometimes also referred to as
quasi-perpendicular kinematics) and the values $q=460$ MeV/c and
$q=995$ MeV/c have been chosen. In each case the value selected of the
transferred energy $\omega$ almost corresponds to the quasi-elastic
peak value, i.e., $\omega=100$ MeV and $439$ MeV, respectively.  FSI
are taken into account for $q=460$ MeV/c through the optical potential
of Comfort and Karp~\cite{Com80}, which is appropriate for proton
kinetic energies below 183 MeV.  In the case of $q=995$ MeV/c, as the
proton kinetic energy is of the order of $\sim 430$ MeV, we use
instead a Sch\"rodinger-equivalent form of the relativistic global
optical potential of ref.~\cite{Coo93}. Thus the non-relativistic wave
functions correspond to the upper components of the relativistic ones,
containing the Darwin term. Results obtained within this approach were
compared with a fully relativistic calculation in the impulse
approximation~\cite{Udi99}, i.e., without including the effects of
MEC.

In each panel of fig.~6 we compare the results corresponding to the
traditional non-relativistic model with non-relativistic kinematics
(denoted by NR), including relativistic kinematics (RK), and finally,
the semi-relativistic approach discussed in the previous section (SR).
All the calculations include MEC. From these results it emerges that,
particularly for high $q$, the relativistic kinematics plays a crucial
role in describing properly the form of the momentum distribution.
Note that the allowable missing momentum values are determined by the
relation $|p'-q| \leq p \leq p'+q$, with $p'$ fixed by energy
conservation. Assuming non-relativistic kinematics the value of $p'$
is given by
\begin{equation}
p' = \sqrt{2m_N(\epsilon_\nh+\omega)} \, ,
\label{NRK}
\end{equation}
while for relativistic kinematics it results
\begin{equation}
p'= \sqrt{(m_N+\epsilon_\nh+\omega)^2-m_N^2}
 =
\sqrt{2m_N(\epsilon_\nh+\omega)
\left(1+\frac{\epsilon_\nh+\omega}{2m_N}\right)} \, .
\label{RelK}
\end{equation}
In eqs.~(\ref{NRK},\ref{RelK}), the energy $\epsilon_\nh$ represents the
(negative) energy of the bound nucleon. Hence in the case of RK we
solve the Schr\"odinger equation with equivalent energy
$(\epsilon_\nh+\omega)\left(1+\frac{\epsilon_\nh+\omega}{2m_N}\right)$ for
the ejected proton instead of the NR energy $\epsilon_\nh+\omega$.

Once relativistic kinematics is selected, the use of SR currents
produces an enhancement of the $L$ and $TL$ responses and a reduction
of the $T$ and $TT$ ones, whose magnitude increases with $q$. These
effects are connected with the factors $\frac{\kappa}{\sqrt{\tau}}>1$
and $\frac{\sqrt{\tau}}{\kappa} \simeq \frac{1}{1+\tau}$ that enter in
the SR expressions of the longitudinal and transverse currents,
respectively~\cite{Rep02,Ama02b}. It is important to point out the
particularly large relativistic enhancement observed for the
interference $TL$ response. This is connected to the spin-orbit
term~\cite{Ama96b} contained in the SR charge density but neglected
within the NR approach. As shown in~\cite{Ama96b}, the interference
between the spin-orbit term and the magnetization current gives rise
to a contribution in $W^{TL}$ which is of the same order of magnitude
as the interference between the charge density and convection
current. Thus the presence of the spin-orbit term in the density is
essential to describe properly the response $W^{TL}$, even for
moderate $q$-values.

Analogous results hold for a proton knockout from the $p_{3/2}$-shell
in $^{16}$O: the role of relativity in each response is similar to the
one observed for the $p_{1/2}$ case and, again, the most sensitive
response to relativistic effects is $W^{TL}$. However, the
relativistic enhancement for the $p_{3/2}$-orbit is somewhat smaller
than the one presented for $p_{1/2}$, particularly for large $q$.

%----------------
%    Figure 7
%----------------
\begin{figure}[tp]
\begin{center}
\leavevmode
% scala la figura di  un fattore 0.9:
\def\epsfsize#1#2{0.8#1}
\epsfbox[50 340 500 730]{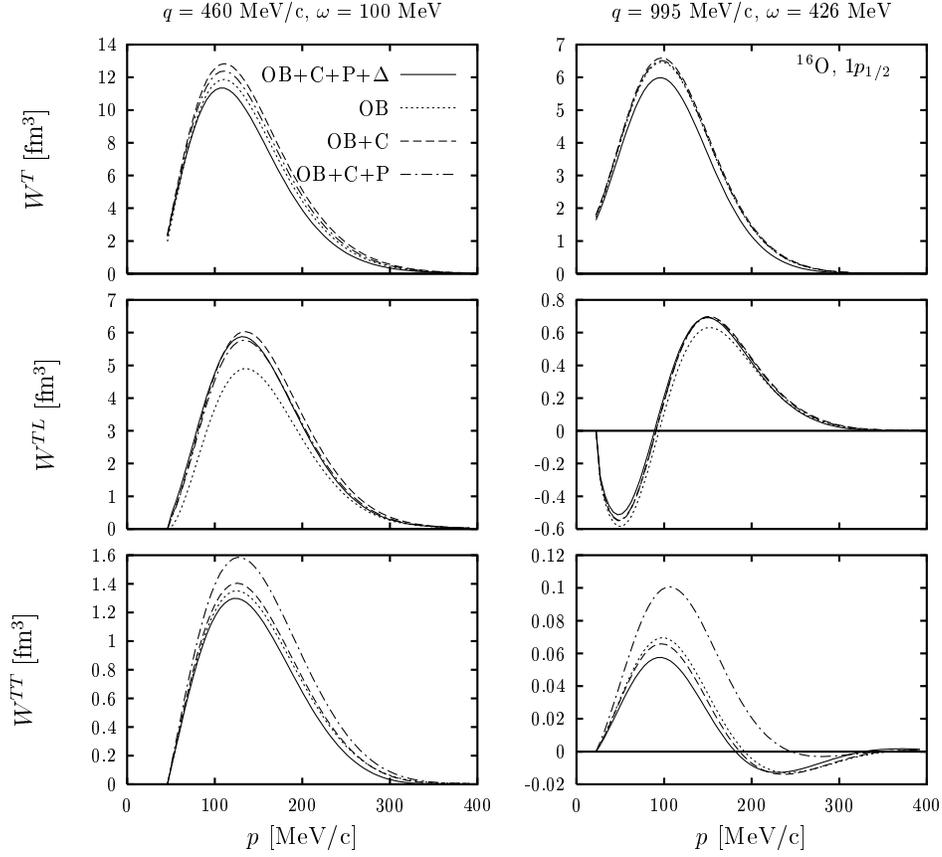}
\end{center}
\caption{ Separate contribution of each one of the MEC to the
exclusive response functions. SR results are shown for knock-out from
the $1p_{1/2}$ shell of $^{16}$O for the same kinematics as in fig.~6,
including the one-body current only (OB), and in addition the contact
(C), pionic (P) and $\Delta$ currents in the calculation.  }
\end{figure}

%----------------
%    Figure 8
%----------------
\begin{figure}[t]
\begin{center}
\leavevmode
% scala la figura di  un fattore 0.9:
\def\epsfsize#1#2{0.8#1}
\epsfbox[50 340 500 730]{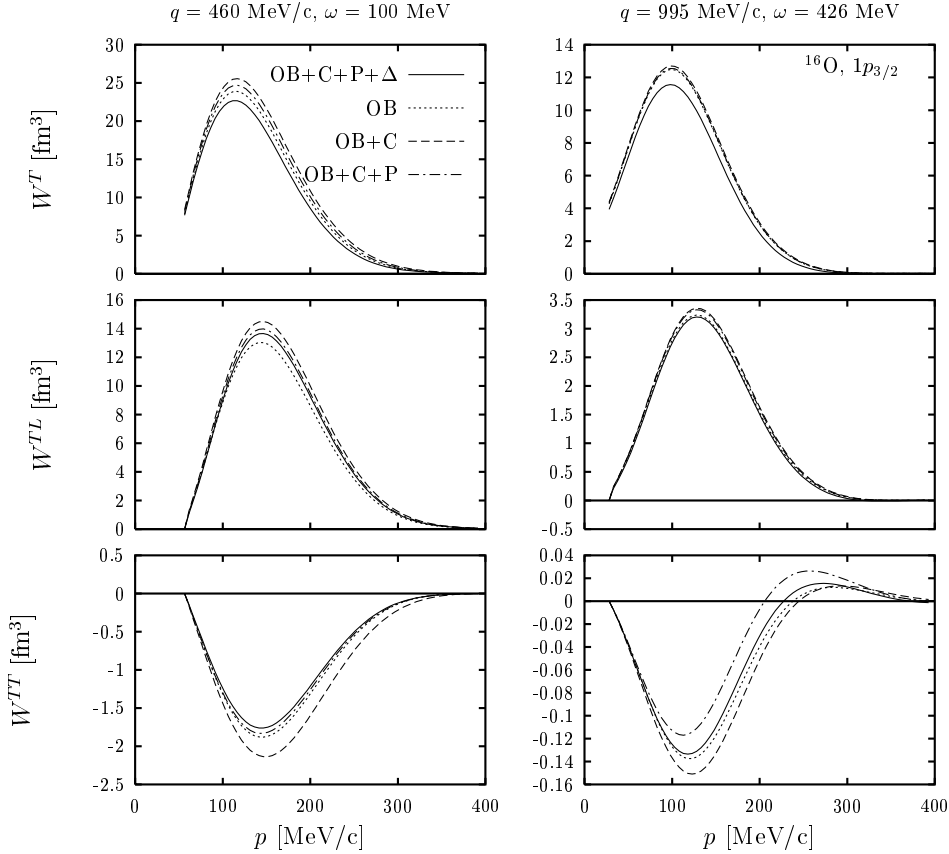}
\end{center}
\caption{
The same as fig.~7 for knock-out from the $1p_{3/2}$ shell of $^{16}$O.
}
\end{figure}

The separate contributions of the MEC is presented in figs.~7 and 8
for $p_{1/2}$ and $p_{3/2}$-shells, respectively. Kinematics is as in
the previous figure and we do not show results for the pure
longitudinal response as it is not affected by MEC within the present
approaches. We compare the OB responses (dotted line) with the results
obtained when including the contact (C) current (dashed line), the
contact and pion-in-flight (C+P) currents (dot-dashed line), and the
contact, pionic and $\Delta$ (C+P+$\Delta$) currents (solid line).
Let us analyze each response separately. In the case of $W^T$, the
contact current produces an increase which is partially cancelled by
the tiny reduction introduced by the pionic current. Note that the
role of the C and P currents is negligible for large $q$.  The
$\Delta$ current gives rise to an additional reduction whose relative
magnitude is rather similar for both shells.  The total effect of
C+P+$\Delta$ MEC is a reduction of the T response which is slightly
more important for high $q$.

Although a great caution should be taken in extending the conclusions
drawn from the analysis of exclusive reactions to inclusive ones or
viceversa, it is illustrative to discuss the results relative to
quasielastic $(e,e'p)$ reactions in connection with the quasielastic
$(e,e')$ ones. Note that, apart from the potentials usually considered
in both types of reactions, the inclusive responses are calculated by
integrating the exclusive ones over the ejected nucleon variables, and
summing over all occupied hole states, i.e., including also the
contribution given by the $1s_{1/2}$ shell.  In
refs.~\cite{Rep02,Ama02b,Ama03} we evaluated the role of MEC for the
inclusive $T$ response in a relativistic Fermi gas model. The contact
(C) contribution was found to be larger than the pion-in-flight term
(P), a dominance increasing with $q$.  The total (C+P) contribution
presents an oscillatory behaviour with respect to the transferred
energy $\omega$, having a node close to the QEP value, particularly
for high $q$. For $q=1$ GeV/c, this node is reached for $\omega\sim
435$ MeV, which is very close to the value $\omega=439$ MeV selected
in the $(e,e'p)$ calculations (right hand panels in figs.~7 and 8).
On the contrary, in the case $q=500$ MeV/c, the contribution of the
C+P currents in the inclusive $T$ response presents a node for
$\omega\sim 150$ MeV and is positive and non-negligible for the value
$\omega=100$ MeV selected here in the case of $(e,e'p)$ process.  A
similar discussion can be also applied to the $\Delta$ results. In
ref.~\cite{Ama03} (see also previous section) the $\Delta$
contribution to the inclusive $T$ response has been shown to be always
negative up to $q\sim 2$ GeV/c and larger than the one provided by the
C and P currents together. Hence the general trend of the MEC
contribution to the inclusive channel agrees with the effects shown
for the exclusive responses (figs.~7 and 8).

Finally, comparing the results in figs.~7 and 8 we conclude that the
effects of MEC for the exclusive $T$-response are similar for both
$p_{1/2}$ and $p_{3/2}$ shells.  A different behaviour is found in the
results of ref.~\cite{Giu02} where the role of MEC, for $q=460$ MeV/,
is shown to be larger in the case of the $p_{3/2}$ shell. In addition,
the role introduced by the pionic current (P) in~\cite{Giu02} is said
to be negligible, whereas in our case its contribution, though a
little bit smaller than the contact term, is clearly visible.  Our
results are also in disagreement with the calculation of~\cite{Ryc99}
performed for $q=1$ GeV, where the total MEC effect is found to be
small for the $p_{1/2}$ and large for the $p_{3/2}$ in the $T$
response.  This discrepancy is linked to the FW method used to
implement relativity in~\cite{Ryc99} which is not expected to be
applicable for $q=1$ GeV~\cite{Ama02b}.

Next we focus on the interference $TL$ response. From figs.~7 and 8 we
observe that for $q=460$ MeV/c the MEC effect is larger in the case of
the $p_{1/2}$-shell. The main contribution comes from the contact
current, which produces an enhancement in $W^{TL}$ of the order of
$\sim 20\%$ for $p_{1/2}$ and $\sim15\%$ for $p_{3/2}$, while the role
of the pion-in-flight is smaller, reducing the response, and in
particular, the $\Delta$ gives rise to an almost negligible
contribution (slightly positive for $p_{1/2}$ and negative for
$p_{3/2}$).  These results disagree with the findings of
ref.~\cite{Giu02}, where the effect of the $\Delta$ in $W^{TL}$ is
substantially large and negative for the $p_{1/2}$-shell, so that it
cancels out the contact term, yielding a negligible global MEC
contribution. On the contrary, in the same reference the $\Delta$
contribution for the $p_{3/2}$-orbit is found to be large and
positive, hence the net effect of MEC is an important enhancement in
the $TL$ response.  As in the case of the pure $T$ response, the
pionic contribution found in~\cite{Giu02} is negligible, which again
is not in accord with our results.

Figs.~7 and 8 clearly show that the importance of the MEC on the $TL$
response decreases as $q$ goes to higher values.  Whereas the contact
current enhances the $TL$ response by the same magnitude for the two
$p$-shells, the $\Delta$ term is negligible for the $p_{1/2}$ and
tends to cancel the contact contribution for $p_{3/2}$. For both
shells, the pionic current does not alter $W^{TL}$.

As far as $W^{TT}$ is concerned, note that this response is much
smaller, its contribution being of order $(k_F/m_N)^2$ (see
ref.~\cite{Ama96b}); hence terms of second order in $p/m_N$, usually
neglected in the expansion of the current operators, may provide a
crucial role in this response. However, since $W^{TT}$ is also
particularly sensitive to the details of the model, it can be used as
a test to compare different theoretical models.  Note that this
response is opposite in sign for $p_{1/2}$ and $p_{3/2}$. Concerning
the role of MEC, it is found to be of the same relative order of
magnitude as in the other responses, except for the pionic current
which has an important effect in this case.

%----------------
%    Figure 9
%----------------
\begin{figure}[t]
\begin{center}
\leavevmode
% scala la figura di  un fattore 0.9:
\def\epsfsize#1#2{0.99#1}
\epsfbox[60 460 505 730]{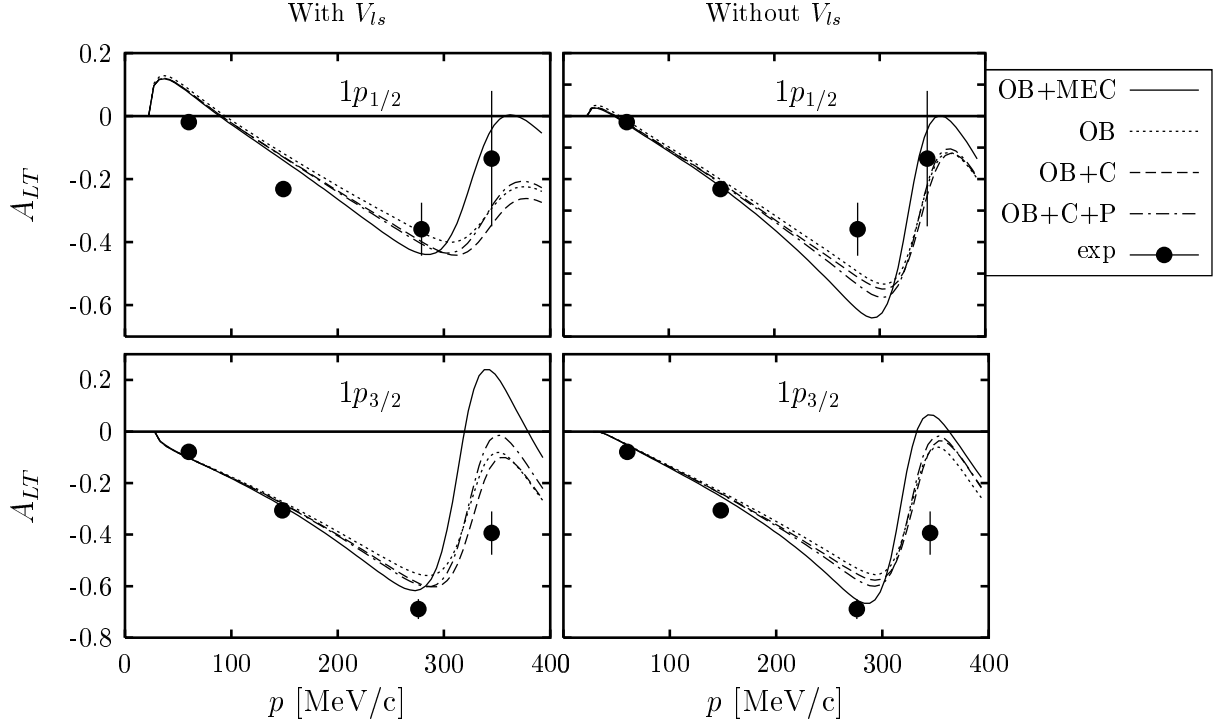}
\end{center}
\caption{$A_{TL}$-asymmetry for proton knock-out from the $1p_{1/2}$
and $1p_{3/2}$ shells in $^{16}$O, computed within the SR model
including the successive contribution of each one of the MEC.
Experimental data are from~\protect\cite{Gao00}.  Left and right
panels have been obtained to including or not the spin orbit part in
the optical potential. }
\end{figure}

As a further application of our model we present in fig.~9 the $TL$
asymmetry ($A_{TL}$) for $p_{1/2}$ (top panels) and $p_{3/2}$ (bottom
panels). The kinematics has been selected to correspond to the
experimental data~\cite{Gao00}. $A_{TL}$ is obtained from the
difference of cross sections measured at $\phi'=0^o$ and $\phi'=180^o$
divided by the sum, hence this observable is particularly interesting
because it does not depend on the spectroscopic factors.  A detailed
study on the $TL$ asymmetry has been carried out in
refs.~\cite{Udi99,Udi01} within the relativistic distorted-wave
impulse approximation (RDWIA), namely a fully relativistic calculation
without including MEC. In particular, a comparison between $A_{TL}$
evaluated in RDWIA and using the SR approach for the one-body current
and neglecting MEC, was presented in~\cite{Udi99}.  There it was
proved the large contribution given by the spin-orbit correction to
the charge density, and, more importantly, the crucial role played by
the dynamical enhancement of the lower components of bound Dirac
spinors in the description of interference $TL$ observables (see also
refs.~\cite{Cab98,Cris02}).  Recent data on polarization observables
nicely agree with RDWIA analysis~\cite{Woo98,Udi00}.

However, a direct calculation of the role of MEC in $A_{TL}$ within a
relativistic approach has never been performed, as far as we know.
Thus, in fig.~9 we show the results obtained within the impulse
approximation, i.e., without MEC (dotted line), and the contributions
introduced by the various two-body currents: C (dashed line), C+P
(dot-dashed line) and total MEC, namely C+P+$\Delta$ (solid line).  In
order to investigate also the dependence of our results on the FSI, in
fig.~9 we show two sets of calculations for each $p$ shell: including
the spin-orbit part of the optical potential $V_{ls}$ \cite{Coo93}
(left panels) and without it (right panels).  For comparison we also
present the experimental data~\cite{Gao00}. From inspection of fig.~9
we conclude that the effect of MEC is very small for low missing
momentum values and it starts to be important for $p\geq 300$
MeV/c. Note however that in this region the dynamical enhancement of
lower components, obviously not considered within the present SR
approach, starts also to play a crucial role and hence it should be
considered before a detailed comparison with data can be
accomplished. Note also the large discrepancy between the SR
calculation and the data for low $p$ in the case of $p_{1/2}$ with the
full potential.  This issue was already present in~\cite{Udi99}, where
the RDWIA calculation differs from the SR one comparing better with
data~\cite{Gao00}.  This problem within the SR approach appears to be
connected with the spin-orbit term introduced by the equivalent
Schr\"odinger form of the optical potential, as can be seen in the
right panels of fig.~9, that do not include $V_{ls}$ in the FSI.  Our
results also show that the MEC effects for high missing momenta
strongly depend upon the FSI, since they are substantially reduced
when only the central part of the optical potential is included (right
panels).

Summarizing, from theoretical results in fig.~9 and ref.~\cite{Udi99},
and their comparison to experimental data~\cite{Gao00}, we may
conclude the following: i) a fully relativistic calculation within the
impulse approximation (RDWIA), i.e., including the effects introduced
by the dynamical enhancement of the lower components in the Dirac
spinors, appears to be essential to reproduce the data; ii) the
effects introduced by MEC for high missing momentum values seem to be
also very important, and highly dependent on the FSI. Hence an
appropriate relativistic analysis of these two-body currents may be
also essential in order to improve the description of the experimental
data at high missing momentum.

%=====================
\section{Conclusions}
%=====================

In this paper we have presented a semi-relativistic model of inclusive
and exclusive electron scattering from nuclei in the one-nucleon
emission channel, including one- and two-body currents.  These
currents differ from the usual non-relativistic ones by
multiplicative $(q,\omega)$-depending factors obtained by an expansion
in powers of the missing momentum. An essential ingredient of the
SR consists in using, in addition to the SR currents, relativistic
kinematics to relate the energy and momentum of the ejected nucleon.

This model has been already tested in quasielastic inclusive $(e,e')$
processes in refs.~\cite{Rep02,Ama02b,Ama03}. In this paper we have
first focussed on the new SR-$\Delta$ current and compared its
contribution to the inclusive transverse response with a fully
relativistic calculation in RFG for moderate and high $q$-values. This
has allowed us to test the reliability of different prescriptions
introduced to account for ``dynamical'' aspects of the $\Delta$
propagator.  We have found that the best agreement with the fully
relativistic calculation corresponds to the static limit
approximation. Thus, we conclude that the use of any of these
``dynamized'' $\Delta$ propagators within the one-particle emission
channel is not justified and, moreover, it may produce very large
discrepancies with the exact result.

Next we have implemented the SR currents into a DWIA model of the
quasielastic $(e,e'p)$ reaction, computing the separate response
functions and the $TL$ asymmetry, using the static limit for the
$\Delta$ propagator.  After analyzing the role of the different
relativistic corrections embedded in our calculation and its
dependence upon the momentum transfer, we have studied the effect of
MEC on the different observables. In particular, we have compared the
results for proton knock-out from the $p_{1/2}$ and $p_{3/2}$ shells
in $^{16}$O, choosing quasi-perpendicular kinematics, typical of the
experiments.  In the case of the $T$ response, MEC effects are shown
to be equally important for the two shells, giving rise to a net
reduction of the response. For $W^{TL}$, the role of MEC is to enhance
the response, an effect that is substantially larger for the $p_{1/2}$
shell, and reduces considerably for high $q$-values. In general, we
get sizable differences with previous calculations in the
literature~\cite{Ryc99,Giu02}.

Finally, our model applied to the $A_{TL}$ asymmetry shows that this
observable is very sensitive to the MEC and to FSI in the region of
high missing momentum. Since it has been proved in~\cite{Udi99} that
the interference $TL$ observables, particularly $A_{TL}$, are also
crucially affected by other relativistic ingredients, such as the
dynamical enhancement of lower components, not included in this work,
it would be very interesting to evaluate MEC effects within the scheme
of the fully relativistic calculation of
refs.~\cite{Udi93,Udi99,Udi01}, and contrast their predictions with
the ones obtained with the present SR model.

%***************************
\section*{Acknowledgments}
%***************************
This work was partially supported by funds provided by DGI (Spain) and
FEDER funds, under Contracts Nos BFM2002-03218, BFM2002-03315 and
FPA2002-04181-C04-04 and
by the Junta de Andaluc\'{\i}a and by the INFN-CICYT exchange.
M.B.B. acknowledges financial support from MEC (Spain) for a
sabbatical stay at University of Sevilla (ref. SAB2001-0025).

%===============================

\end{document}